\documentclass{article}
\usepackage[english]{babel}
\usepackage{amsfonts,amssymb, amsmath}

\newtheorem{prop}{Proposition}

\def\nn{\nonumber }
\def\bq{ \begin{equation}}
\def\eq{ \end{equation}}
\def\ben{ \begin{eqnarray}}
\def\en{ \end{eqnarray}}

\begin{document}


\title{On tensor invariants for integrable cases of Euler, Lagrange and Kovalevskaya rigid body motion}
\author{A.V. Tsiganov\\
\it\small St.Petersburg State University, St.Petersburg, Russia\\
\it\small e--mail: andrey.tsiganov@gmail.com}

\date{}
\maketitle

\begin{abstract}
We discuss global tensor invariants of a rigid body  motion  in the cases of Euler, Lagrange and Kovalevskaya.
These invariants are obtained by substituting tensor fields with cubic on variable components into the invariance equation and solving the resulting algebraic equations using computer algebra systems. According to the Poincar\'{e}-Cartan theory of invariants, an existence of invariant geometric structures raises the question of using them to study the dynamics. 
\end{abstract}

\section{Introduction}
\setcounter{equation}{0}
Let us consider an autonomous system of ordinary differential equations 
\bq\label{m-eq}
\frac{d}{dt} {x} = X(x^1,\ldots,x^n)\,,
\eq 
on the phase space  with coordinates $x=(x^1,\ldots,x^n)$. Tensor field $T$ is an invariant of these equations if it is a constant
\[T(x)=c\]
along any solution of system (\ref{m-eq}).  For sufficiently smooth vector field $X$ 
tensor invariants  $T$ of the flow generated by  $X$ satisfy equation 
\bq\label{g-inv}
\mathcal L_X\,  T=0\,,
\eq 
see  \cite{koz92,koz13,koz14,koz19,rat24} and references therein  for a modern discussion of the relationship between equations (\ref{m-eq}) and (\ref{g-inv}). Here, $L_X T$ is a Lie derivative of tensor field $T$ along vector field $X$. 

The Lie derivative $L_X T$ determines the rate of change of the tensor field $T$ under the phase space deformation defined by the flow of the system (\ref{m-eq}). In local coordinates the Lie derivative of the tensor field $T$ of type $(p, q)$ is equal to
\begin{align}
({\mathcal {L}}_{X}T)^{i_{1}\ldots i_{p}}_{j_{1}\ldots j_{q}}=\sum_{k=1}^n X^{k}(\partial _{k}T^{i_{1}\ldots i_{p}}_{j_{1}\ldots j_{q}})
&-\sum_{\ell=1}^n (\partial _{\ell}X^{i_{1}})T^{\ell i_{2}\ldots i_{p}}_{j_{1}\ldots j_{s}}-\ldots - \sum_{\ell=1}^n (\partial _{\ell}X^{i_{p}})T^{i_{1}\ldots i_{p-1}\ell}_{j_{1}\ldots j_{s}}\label{lie-d}\\
&+\sum_{m=1}^n(\partial _{j_{1}}X^{m})T^{i_{1}\ldots i_{p}}_{m j_{2}\ldots j_{q}}+\ldots +\sum_{m=1}^n(\partial _{j_{q}}X^{m})T^{i_{1}\ldots i_{p}}_{j_{1}\ldots j_{q-1}m}=0
\nonumber
\end{align}
where $\partial _s= {\partial }/{\partial x^s}$ is the partial derivative on the $x^s$ coordinate \cite{nt05}.

The Lie derivative commutes with the exterior differentiation operation and satisfies the Leibniz rule. It allows us to construct tensor invariants from a set of basic invariant tensor fields which either have a simpler functional dependence on the variables $x$, or have some special properties or physical interpretation. As a standard example, let us take the Hamiltonian vector fields on the symplectic manifold $\mathbb R^{2n}$
\bq\label{g-ham}
X=PdH\,,
\eq
where $P$ is an invariant Poisson bivector independent on $x$. The second factor $dH$ in the tensor product (\ref{g-ham}) is an invariant 1-form constructed by differentiating the Hamilton function $H$, i.e., the scalar invariant of the equation (\ref{g-inv}), which usually coincides with the mechanical energy of the dynamical system (\ref{m-eq}) \cite{jac-book, app28,w04}.

Following Jacobi \cite{jac-book}, we usually impose an additional constraint  on an invariant tensor field $P$ in (\ref{g-ham})
\bq\label{jac-cond} [\![P,P]\!]=0\,,\eq
which ensures that
Poisson bracket of integrals of motion $f_1$ and $f_2$ is again an integral of motion
\[f_3=Pdf_1\wedge df_2\quad \Rightarrow\quad \mathcal L_X f_3=0.\]
Here [\![X,Y]\!] is the Schouten-Nijenhuis bracket of tensor fields $X$ and $Y$ \cite{nt05}.

In the Poincar\'{e}-Cartan theory of invariants \cite{poi,car} the Jacobi condition (\ref{jac-cond}) is redundant  because tensor product of invariants are also invariant 
\[
\mathcal L_X\left( Pdf_1\wedge df_2\right)=\left(\mathcal L_X P\right)df_1\wedge df_2+Pd\left(\mathcal L_X f_1\right)\wedge df_2+
 Pdf_1\wedge d\left(\mathcal L_X f_2\right)=0\,.
\]
We need to learn how to extract information about solutions of ODE's (\ref{m-eq}) from all the existing tensor invariants satisfying invariance equation $\mathcal L_XT=0$ (\ref{g-inv}).

The properties of the Lie derivative (\ref{lie-d}) do not depend on the properties of the tensor fields to which it is applied, i.e., it does not depend on symmetry or antisymmetry at index permutation or on the equality to zero of the Schouten-Nijenhuis bracket 
\[[\![T,T]\!]=0,\]
 which is equivalent to fulfilling the Jacobi identity for Poisson brackets and its analogues \cite{s40,n55,fil,taht}.  Remind, that for the outer product of vector fields, the Schouten-Nijenhuis bracket is defined using the Lie bracket $[.,.]$
\begin{align}\label{br-sn}
 [\![X_1\wedge\dots \wedge X_k, Y_1\wedge \dots
\wedge Y_\ell]\!] = \sum_{i,j}(-1)^{i+j} [X_i,Y_j] \wedge X_1\wedge\\
\dots \wedge \widehat{X}_i \wedge \dots \wedge X_k\wedge
Y_1\wedge \dots \wedge \widehat{Y}_j\wedge \dots \wedge Y_\ell\,.\nn
\end{align}
Here $\widehat{X}_i$ and $\widehat{Y}_j$ mean that the corresponding vector fields ${X}_i$ and ${Y}_j$ are missing in the outer product of the vector fields \cite{mm95}.

When the divergence of the vector field $X$ is zero
\bq\label{z-div}
\mbox{div} X=\partial_1 X^1+\partial_2 X^2+\cdots+\partial_n X^n=0\,,
\eq
then the Lie derivative (\ref{lie-d}) of completely antisymmetric unit tensor fields $\Omega$ of type $(0,n)$ and $\mathcal E$ of type $(n,0)$ is zero
\bq\label{vol-inv}
 \Omega =dx^1\wedge dx^2\wedge\cdots\wedge x^n\,,
\qquad \mathcal L_X\Omega=0\,,
\eq
and
\bq\label{lc-t}
 \mathcal E= \partial_1\wedge\partial_2\wedge\cdots\wedge\partial_n\,,\qquad
 \mathcal L_X \mathcal E=0\,.
\eq
According to \cite{taht}, in generic case invariant tensor $\mathcal E$ does not satisfy to the Jacobi identity, but it doesn't play any role in the theory of invariants. 

The study of the invariance of the differential form $\Omega$ begins in 1838 with the Liouville theorem \cite{liu38} on the conservation of the volume form for the vector fields without divergency
\bq
\int_V \Omega=const\,,
\eq
and continues in the Poincar\'{e}-Cartan theory of invariants \cite{poi,car}. The invariance of the multivector field $\mathcal E$ is a key element of the Jacobi theory of functional determinants \cite{jac0,jac1,jac2,jac-book0} developed in 1841-1845.

The invariant tensor fields $\Omega$ and $\mathcal E$ give rise to a tower of the tensor invariants if we know first integrals $f_1,\ldots,f_k$ of equations (\ref{m-eq}). For instance tensor fields
\bq\label{tow-t}
T=\mathcal E\,,\qquad T_i=\mathcal E df_i\,,\qquad T_{ij}=\mathcal E df_idf_j\,,\qquad T_{ijk}=\mathcal Edf_1df_jdf_k\,,\ldots,
\eq
are invariant tensor fields  for the flow of any divergence free vector field $X$. 

 We emphasise that we are talking about global tensor invariants uniquely defined on the whole phase space, since locally, for example in the neighbourhood of Liouville tori for Kol\-mo\-go\-rov non\--de\-ge\-ne\-rate systems, all solutions of the invariance equation (\ref{g-inv}) in spaces of differential 2-forms and Poisson bivectors can be constructed using action-angle variables \cite{bog96}.

In this paper we will discuss solutions of the invariance equation (\ref{g-inv}) for the equations of motion describing the rotation of a rigid body for three well-known integrable cases - Euler, Lagrange and Kovalevskaya, the description of which can be found in the book \cite{bm}.  In all these cases the divergence of the vector field is zero and to find the tensor invariants we will solve the invariance equation (\ref{g-inv}) using polynomial substitution on the variables $x_1,\ldots,x_n$ for the components of the tensor field $T$ of the form
\bq\label{cub-anz}
T^{i_{1}\ldots i_{p}}_{j_{1}\ldots j_{q}}=\sum_{k,\ell,m=1}^n (a^{i_{1}\ldots i_{p}}_{j_{1}\ldots j_{q}})^{k\ell m}x_k x_\ell x_m
+\sum_{k,\ell}^n (b^{i_{1}\ldots i_{p}}_{j_{1}\ldots j_{q}})^{k\ell}x_k x_\ell 
+\sum_{k=1}^n (c^{i_{1}\ldots i_{p}}_{j_{1}\ldots j_{q}})^{k}x_k 
+d^{i_{1}\ldots i_{p}}_{j_{1}\ldots j_{q}}\,.
\eq
Here $a,b,c$ and $d$ with the corresponding indices are unknown real parameters, which are determined in the process of solving the system of algebraic equations resulting from the substitution of (\ref{cub-anz}) into the invariance equation (\ref{g-inv}).

We use various computer algebra systems to find, classify and verify solutions. According to \cite{arn06}, computer computation allows us to find new solutions and to draw attention to the necessary analytical study of the properties of computer-generated solutions.

The main result of this mathematical experiment is the fact that for Lagrange and Kovalevskaya systems all solutions of the invariance equation (\ref{g-inv}) in the space of differential forms obtained by computer methods can be constructed from  $\Omega$ and first integrals by means of differentiation and tensor multiplication, while in the space of multivector fields there exist solutions which cannot be obtained from the basic invariant $\mathcal E$ and first integrals by means of the same operations.

\subsection{The Jacobi theory of functional determinants}
In modern literature, the term "Jacobian" is often used interchangeably to refer to both the Jacobian matrix and its determinant. 
The Jacobian matrix collects all first-order partial derivatives of a multivariate function, which can be used, for example, for the backpropagation algorithm in training a neural network. The Jacobian determinant or functional determinant is useful in changing between variables, where it acts as a scaling factor between one coordinate space and another. 

Of course, Jacobi was not the first to study the functional determinant which now bears his name, it first appears in a paper by Cauchy in 1815, see historical remarks in \cite{miur}. The name "Jacobian" for functional determinants was proposed in 1853 by Sylvester, see page 476 in \cite{syl53} referring to the Jacobi paper \cite{jac0} published in 1841. In this paper Jacobi proved, among many other things, that if a set of $n$ functions in $n$ variables are functionally related then the Jacobian is identically zero, while if the functions are independent the Jacobian cannot be identically zero.

A few years later, in 1844–45, Jacobi published a substantial memoirs \cite{jac1,jac2} on the
method of the multiplier, which extends to systems of ordinary differential equations
the method of the integrating factor for scalar equations proposed by Euler in \cite{eul}. 
In his lectures on dynamics delivered in 1842-43, published posthumously in 1866,
reprinted (slightly revised) in 1884 as a supplement to his Gesammelte Werke
\cite{jac-book0} and translated into English in \cite{jac-book}, Jacobi developed in Lecture 13 the theory of functional
determinants, essentially following \cite{jac0}, and, in Chapter 14, the theory of the
multiplier. On this occasion, in \cite{jac-book0}, pp. 104-106, or in \cite{jac-book}, pp. 112-114, he again proved the main
properties of Jacobian without isolating they as  separate statements.

The Jacobi theory rapidly became a classical tool, for instance see 
 textbooks on analytical mechanics by Appel \cite{app28} and Whittaker \cite{w04}, the course of lectures for engineers by Vallée-Poussin \cite{vp12}, the course of mathematical analysis by Goursat \cite{gur} and by Bianchi course on differential geometry \cite{b03}. The first editions of these textbooks were published at almost the same time at the beginning of the 19th century.
 Instead of the original Jacobi notation for functional determinants
 \bq\label{jac-det}
Q=\sum \pm \frac{\partial g_1}{\partial x_1}\frac{\partial g_2}{\partial x_2}\cdots\frac{\partial g_n}{\partial x_n}\,,
\eq
these textbooks use the notation 
\[
Q=\frac{\partial (g_1,g_2,\ldots, g_n)}{\partial (x_1,x_2,\ldots,x_n)}\,,
\]
which we will  use below.

Let us consider a system of differential equations using notation from the Jacobi paper \cite{jac2} with the lower indices: 
\bq\label{d-eq} \frac{dx_1}{X_1}=\frac{dx_2}{X_2}=\cdots=\frac{dx_n}{X_n}\,, 
\eq 
where $X_i$ are functions on phase space with coordinates $x_1,\ldots,x_n$, see discussion of these equations in the theory of invariants \cite{poi,car}.

Any function $f(x_1,\ldots,x_n)$ that is constant at $x_1,\ldots,x_n$ satisfying (\ref{d-eq}) is the first integral and this  condition Jacobi  writes in the following form "for the sake of brevity"
\bq\label{def-int}
X(f)=\frac{\partial f}{\partial x_1}X_1+\frac{\partial f}{\partial x_2}X_2+\ldots+\frac{\partial f}{\partial x_n}X_n=0\,,
\eq
Of course, it is a Lie derivative of $f$ along vector field $X$ 
\[\mathcal L_X f=0.\]
The system (\ref{d-eq}) can only have $n-1$ independent first integrals $f_1,\ldots,f_{n-1}$, and any other integral $f$ must be a function of $f_1,\ldots,f_{n-1}$ because the functional determinant is zero
\[
\frac{\partial(f,f_1,\ldots,f_{n-1})}{\partial(x_1,\ldots,x_n)}=0\,.
\]
Decomposing this determinant by the elements of the first row we obtain
\bq\label{jac-1}
\frac{\partial f}{\partial x_1}\Delta_1+\frac{\partial f}{\partial x_2}\Delta_2+\cdots+
\frac{\partial f}{\partial x_n}\Delta_n=0\,,
\eq
where $\Delta_i$ is the algebraic complement of the $i$-th element on the first row of the Jacobian. At $X(x)\neq 0$, the coefficients on the derivatives in equations (\ref{def-int}) and (\ref{jac-1}) are proportional, i.e., there exists a function $M(x)$ such that
\bq\label{m-def}
M(x)X_i=\Delta_i\,.
\eq
The function $M(x)$ is called the multiplier of a system of equations (\ref{d-eq}) that satisfies a linear partial differential equation
\bq\label{g-mul}
\mbox{div} (MX)=\partial_1(MX_1)+\partial_2 (MX_2)+\cdots+\partial_n(MX_n)=0\,,
\eq
in which the first integrals $f_1,\ldots,f_{n-1}$ are missing.  The paramount importance of the Jacobi multiplier $M$ derives from the fact that if $M$ and $M'$ are solutions of (\ref{g-mul}), then the quotient $M/M'$ is a first integral of the dynamical system (\ref{m-eq}). Relations between these first integrals and symmetries are discussed in the Bianchi textbook \cite{b03}, see also \cite{n05}.

The Jacobi definition (\ref{m-def}) can be rewritten as follows
\[
MX=\frac{\partial(x,f_1,\ldots,f_{n-1})}{\partial(x_1,\ldots,x_n)}\,,
\]
for example, see page 253 of the Vall\'{e}e-Poussin textbook \cite{vp12} and discussion in the Kozlov papers \cite{koz19,biz15}.

Thus, Jacobi  obtained representation of the vector field $X$ through the functional determinant and  multiplier
\bq\label{wp}
X=\frac{1}{M(x)}\frac{\partial(x,f_1,\ldots,f_{n-1})}{\partial(x_1,\ldots,x_n)}\,.
\eq
Jacobi also proved that if the divergence of the field $X$ is zero (\ref{z-div}), then the multiplier $M(x)$ is a function of the first integrals, which can always be chosen so that $M(x)=1$ and
\[
X=\frac{\partial(x,f_1,\ldots,f_{n-1})}{\partial(x_1,\ldots,x_n)}\,.
\]
Inversely, if the dynamical system is representable through the Jacobians of the first integrals, then it is divergence-free, see the corresponding theorems in section 229 of the Vall\'{e}e-Poussin lectures for engineers \cite{vp12}. 

Definition  (\ref{jac-det}) includes a completely antisymmetric unit tensor $\mathcal E$ and, therefore, the vector field $X$ can be rewritten as
\bq\label{wp2}
X=\frac{1}{M(x)}\,\mathcal E\, df_1\cdots df_{n-1}=T_n\omega_{n-1}
\eq
where 
\[
T_n=\frac{1}{M(x)}\,\mathcal E \qquad\mbox{and}\qquad \omega_{n-1}=df_1\cdots df_{n-1}
\]
are the tensor invariant of $X$, i.e. $\mathcal L_X T_n=0$ and $\mathcal L_X \omega_{n-1}=0$.

To prove these results, Jacobi uses the fundamental lemma given on page 203 in the first part of the \cite{jac2}, written in Latin, according to which the algebraic complements $\Delta_{ij}$ in the Jacobian of the mapping $g:\mathbb R^n\to \mathbb R^n$ satisfy the relations
\bq\label{2-eq}
\sum_{j=1}^n \frac{\partial}{\partial x_j}\,\Delta_{ij}=0\,, 
\eq
so that
\bq\label{3-eq}
\frac{\partial(g_1,g_2,\ldots,g_{n})}{\partial(x_1,\ldots,x_n)}=\sum_{j}^n \left( \frac{\partial g_i}{\partial x_j}\right)\,\Delta_{ij}
=\sum_{j}^n \frac{\partial }{\partial x_j}\,\bigl(g_i\Delta_{ij}\bigr)\,.
\eq
In the Appel textbook \cite{app28}, which according to his contemporary Bobylev was the best course on theoretical mechanics of its time, the Jacobi proof of this lemma is reproduced on pages 460-461.

The Jacobi theory of  functional determinants is applied not only in the theory of ordinary differential equations and analytical mechanics. For example, in 1910 Adamar, in an appendix to the Tannery textbook on real functions \cite{had10}, introduced the notion of Kronecker index for continuous and non vanishing mappings on smooth boundary using a generalisation of the equations (\ref{2-eq}) and (\ref{3-eq}). The same formulae are used in the definition of the Brouwer degree in \cite{h58}, in the Brouwer fixed point theorem \cite{du59}, in the theory of quasi convex functions \cite{mor66}, in the theory of 
Jacobian in the sense of distributions \cite{resh68}, in the theory of null Lagrangians \cite{olver88}, etc.

\subsection{Tensor invariants for system with $n-1$ first integrals}
In the \cite{jac-book} Jacobi treats the theory of functional determinants as an introduction to the construction and study of Hamiltonian systems, see the last paragraph of Lecture 18. 

For vector fields without divergence one can set $M(x)=1$ and represent the functional determinant  (\ref{wp}-\ref{wp2}) in various ways as a combination of elements of the Jacobian and their algebraic complements. The tower of brackets corresponding to these combinations for arbitrary functions on the phase space $g_1,\ldots,g_m$, $2\leq m\leq n$, looks like
\begin{align}
\{g_1,g_2,\ldots,g_n\}=&\frac{\partial(g_1,g_2,\ldots,g_{n})}{\partial(x^1,x^2,\ldots,x^n)}=\mathcal E dg_1\cdots dg_{n}\,,\nn\\
\nn\\
\{g_1,\ldots,g_{n-1}\}_{\alpha_1}=
&\frac{\partial(g_1,\ldots,g_{n-1}, f_{\alpha_1})}{\partial(x^1,x^2,\ldots,x^n)}=T_{\alpha_1} dg_1\cdots dg_{n-1}\,,\nn\\
\nn\\
\{g_1,\ldots,g_{n-2}\}_{\alpha_1\alpha_2}=
&\frac{\partial(g_1,\ldots,g_{n-2},f_{\alpha_1},f_{\alpha_2})}{\partial(x^1,x^2,\ldots,x^n)}
=T_{\alpha_1,\alpha_2} dg_1 \cdots  dg_{n-2}\,,\label{inv-chain}\\
\vdots\nn\\
\{g_1,g_2\}_{\alpha_1\cdots\alpha_{n-2}}=&\frac{\partial(g_1,g_{2},f_{\alpha_1},\ldots,f_{\alpha_{n-2}} )}{\partial(x^1,x^2,\ldots,x^n)}=T_{\alpha_1,\ldots,\alpha_{n-2}} dg_1 dg_2\,.\nn
\end{align}
If $[\![\mathcal E,\mathcal E]\!]=0$, all these brackets satisfy the Jacobi identity as well as their linear combinations, which is a consequence of the fundamental Jacobi lemma. 

At $m=2$ we have standard Poisson brackets for two functions, and so we can say that the vector field $X$ (\ref{wp2}) is a Hamiltonian vector field (\ref{g-ham}).
\begin{prop}
If $[\![\mathcal E,\mathcal E]\!]=0$, a divergence-free vector field with $n-1$ independent first integrals (\ref{z-div}) is Ha\-mil\-to\-nian with respect to each of its first integrals
\[
X=\{x,f_k\}_{1\cdots\hat{k}\cdots (n-1)}  = P_k\,df_k\,,\qquad k=1,\ldots,n-1\,.
\]
The corresponding Poisson invariant bivectors have the form
\bq\label{gen-poi}
P_k=(-1)^{n-k-1} \mathcal E\, df_1\dots  \widehat{df}_k \dots df_{n-1}\,,\qquad k=1,\ldots,n-1\,.
\eq
\end{prop} 
The proof is  a straightforward computation.

To prove the Jacobi identity, we use the fundamental Jacobi lemma and the expression for the components of the  bivector $P_k$ by the algebraic complements $\Delta^{ij}_k$ of the Jacobian elements of dimension $(n-2)\times(n-2)$
\[
P_k^{ij}=(-1)^{i+j}\Delta^{ij}_k\,,\qquad 
\Delta^{ij}_k=\frac{\partial(\hat{x},f_1,\ldots,\hat{f}_k,\ldots f_{n-1})}{\partial(x^1,\ldots,\hat{x}^i,\ldots,\hat{x}^j,\ldots,x^n)}\,,
\]
which are obtained by subtracting from the Jacobian (\ref{wp}) the first and $k$-th row and $i$-th and $j$-th column. On the other hand, since
\[
\{g_1,g_2\}_k=\{g_1,g_2,f_1,\ldots,\hat{f}_k,\ldots,f_{n-1}\,,\}
\]
then we can say that the fulfilment of the Jacobi identity follows from the equality to zero of the Schouten-Nijenhuis bracket for the basic tensor invariant $\mathcal E$. The same Poisson brackets could be obtained using the second basic invariant $\Omega$ \cite{dam12}.

In classical mechanics, examples of the use of the Jacobi theory of functional determinants are the Euler equations describing the rotation of a rigid body \cite{jac4}, the problem of the attraction of a point to a fixed centre in a resistive medium and in empty space \cite{jac-book}, the Jacobi system of three bodies attracted by a force proportional to the cube of the inverse of the distance between them (partial case of the Calogero-Moser system) \cite{jac3}, and other dynamical systems considered by Jacobi himself.  

We should also mention the Volterra series of 1895 \cite{vol99} papers devoted to the study of the Chandler oscillation of the Earth poles, in which the representation of the vector field by the functional determinant (\ref{wp}) is used to study the Euler equations of a rigid body with a gyrostat, and the Jacobi lectures \cite{jac-book0,jac-book}, in which, as an exercise in multiplier theory, additional first integrals are found in the Kepler problem, called  components of the Runge-Lenz vector in physics. For homogeneous dynamical systems with quadratic integrals representation of the vector field for via Jacobian of the first integrals was obtained in  \cite{vol97}. These Volterra results was repeated in \cite{ob81}, Addendum 1, see also \cite{biz15}.

Representation of a vector field as a functional determinant
\[
X=\frac{ \partial(x,f_1,\ldots, f_{n-1}) }{ \partial(x^1,\ldots,x^n) }
\]
for the Euler system without a gyrostat was rediscovered by Nambu \cite{nam73}, which, after the work of Takhtajian \cite{taht}, led to the creation of generalised Nambu mechanics. The Jacobi results for the Kepler problem were rediscovered in \cite{chat96}, see also paper \cite{nutku} about Poisson structures of dynamical systems with three degrees of freedom. A discussion of Poisson-Nambu geometry and related references can be found, for example, in \cite{vai99,grab00,rnam24}.

Similarly to the Kepler problem and the Jacobi three-body problem (partial case of the Calogero-Moser system), the Jacobi representation of the vector field (\ref{wp}-\ref{wp2}) exists for all maximally degenerate or superintegrable Hamiltonian systems on a symplectic manifold of dimension $n=2m$, since the divergence of the Hamiltonian field is zero and maximally degenerate systems have the number of $n-1$ independent first integrals necessary for this representation. There are many examples of such degenerate systems, see examples and references in \cite{evans90,ts08s,ts19s,ts20s}.

A generalisation of the representation (\ref{wp}) to the more general case when the dynamical system  (\ref{m-eq}) admits $n -1$ absolute integral invariants is discussed in the Kozlov paper \cite{koz19}, where the equations of motion for the Goryachev-Chaplygin top are considered as one of the examples.

\subsection{Tensor invariants for system with $n-2$ first integrals}
According to the Euler-Jacobi theorem  equations of motion (\ref{m-eq}) are integrable by quadratures if the vector field $X$ preserves the volume form
\bq\label{gen-mult}
\Omega=M(x) dx^1\wedge dx^2\wedge\cdots\wedge dx^n
\eq
and $n-2$ independent first integrals $f_1,\ldots, f_{n-2}$. Modern discussion of the Euler-Jacobi integrability theorem and its generalisations can be found in \cite{koz13,koz14,koz19}. 

The last multiplier $M(x)$ in (\ref{gen-mult}) satisfies the equation (\ref{g-mul}), but is no longer the first integral of the system of equations (\ref{d-eq}). In \cite{jac1,jac2} Jacobi reduces this case to the previous case with $n-1$ first integrals by constructing the $(n-1)$-th independent first integral
\[f_{n-1}=\int M(y)(Y^2dy^1-Y^1dy^2)\,,\] 
where $y^{1,2}$ and $Y^{1,2}$ are coordinates and components of the vector field in the neighbourhood of the invariant manifold given by the relations $f_1=c_1,\dots,f_{n-2}=c_{n-2}$. 

It is quite natural to ask about the representation of the vector field $X$ satisfying the conditions of the Euler-Jacobi theorem in a form analogous to (\ref{wp2}) 
\bq\label{wp3}
 X=T_m\,\omega_{m-1}\,,
\eq
so that 
\[
L_X X=\left(L_XT_m\right)\omega_{m-1}+T_m\left(\mathcal L_X\omega_{m-1}\right)=0\,. 
\] 
Here $T_m$ is a multivector field of type $(m,0)$ with components that do not depend on the coordinates $x$, and $\omega_{m-1}$ is an invariant differential form of type $(0,m-1)$. Furthermore, we prove that in both the Lagrange and Kovalevskaya cases the invariance condition (\ref{g-inv})
can actually be replaced by a weaker one
\bq\label{ug-inv2}
\left(\mathcal L_X\,T_m\right)\omega_{m-1}=0\,,\qquad \mathcal L_X\omega_{m-1}=0\,.
\eq
Following the terminology of the Poincar\'{e}-Cartan theory of invariants \cite{poi,car}, we will call solutions $T$ of the invariance equation (\ref{g-inv}) absolute tensor invariants and solutions $T_m$ of the equation (\ref{ug-inv2}) conditional tensor invariants depending on absolute invariant $\omega_{m-1}$.

For a vector field of the form (\ref{wp3}), one can construct a counterpart of tower of invariants (\ref{tow-t}) and invariant brackets (\ref{inv-chain}). So, construction of tensor invariants is closely related to the problem of Hamiltonization vector fields possessing an invariant measure $M(x)$.  Recall that the Poisson brackets for non-Hamiltonian dynamical systems integrable by the Euler-Jacobi theorem usually depend on the phase space variables in a rather complicated way \cite{leon99,bbm11,ts12,bmt14,bbm16,ts19}. We can assume that the corresponding Poisson bivectors can be represented as a tensor product of numerical tensor fields $T_m$ and invariant $\omega_{m-1}$ form defined by first integrals $f_i$.

\section{The Euler case}
Let us start with the Euler equations describing the motion of a rigid body
\[
\dot{M}_1=(a_2 - a_3)M_2M_3\,,\qquad \dot{M}_2=(a_3 - a_1)M_3M_1\,,\qquad \dot{M}_3=(a_1 - a_2)M_1M_2\,,
\]
where $M_1,M_2$ and $M_3$ are the coordinates on the phase space of this autonomous system of differential equations (\ref{m-eq}), $a_1,a_2$ and $a_3$ are arbitrary real parameters \cite{bm}. Decomposition of the corresponding vector field $X$ by the basis of the vector fields
\[\partial_i=\frac{\partial}{\partial M_i}\,,\qquad i=1,2,3\,,\]   
looks like
\bq\label{X-eul}
X=b_1M_2M_3\,\partial_1+b_2M_3M_1\,\partial_2+b_3M_1M_2\,\partial_3\,,
\eq
where
\[
b_1=a_2-a_3\,,\qquad b_2=a_3-a_1\,,\qquad b_3=a_1-a_2.
\]

According to \cite{bm}, we choose polynomials of the second degree on the variables $M_k$ as independent scalar invariants
\[f_1=\frac12\left(a_1M_1^2+a_2M_2^2+a_3M_3^2\right)\,,\qquad f_2=M_1^2+M_2^2+M_3^2\,.\]
Any function of these first integrals $g(f_1,f_2)$ is also a solution of the invariance equation (\ref{g-inv}), as are the differential forms of type $(0,1)$ and $(0,2)$ generated by the first integrals
\bq\label{df-1}
\omega_1=dg(f_1,f_2)\,,\qquad \omega_2=dg_1(f_1,f_2) dg_2(f_1,f_2)\,,
\eq
Here $g_{1,2}(f_1,f_2)$ are arbitrary functions of the first integrals, and the tensor product can be symmetrized or alternated, see the discussion in \cite{koz19}.  The degenerate differential forms thus obtained can be used analogously to their application in vortex theory \cite{poi,car,koz13a}.

The divergence of the vector field $X$ is zero and, therefore, in the space of differential forms of type $(0,3)$ the basic solution of the invariance equation (\ref{g-inv}) is a volume form
\[
\Omega=dM_1\wedge dM_2\wedge dM_3\,,
\]
 and the basic solution in the space of multivector fields  of type (3,0) has the form
\bq\label{lcv-eul}
\mathcal E =\partial_1\wedge \partial_2 \wedge \partial_3\,.
\eq
The Schouten-Nijenhuis bracket (\ref{br-sn}) is zero $[\![\mathcal E,\mathcal E]\!]=0$ and, therefore, this trivector defines a Poisson bracket 
\[
\{g_1,g_2,g_3\}=\sum_{i,j,k=1}^3 \mathcal E^{ijk}\frac{\partial g_1}{\partial M_i}  \frac{\partial g_2}{\partial M_j}  \frac{\partial g_3}{\partial M_k}
\] 
so that
\[
X=\{x,f_1,f_2\}\,.
\]
The products of the invariant  trivector $\mathcal E$ and the invariant 1-forms $df_1$ and $df_2$ are compatible Poisson bivectors on the Lie-Poisson algebra $so^*(3)$ 
\[
P_1=\mathcal E df_1=\left(
                      \begin{array}{ccc}
                        0 & a_3M_3 & -a_2M_2 \\
                        -a_3M_3 & 0 & a_1M_1 \\
                        a_2M_2 & -a_1M_1 & 0 \\
                      \end{array}
                    \right)\qquad\mbox{and}\qquad
 P_2=\mathcal E df_2=\left(
                      \begin{array}{ccc}
                        0 & M_3 & -M_2 \\
                        -M_3 & 0 & M_1 \\
                        M_2 & -M_1 & 0 \\
                      \end{array}
                    \right)                   
\]
so that
\[
X=P_1df_2=\mathcal E df_1\wedge df_2\,,\qquad P_1df_1=\mathcal E df_1\wedge df_1=0\,,
\]
and
\[
X=-P_2df_1=-\mathcal E df_2\wedge df_1\,,\qquad P_2df_2=\mathcal E df_2\wedge df_2=0\,.
\]
Remind that two Poisson bivectors $P_1$ and $P_2$ are compatible if their linear combination $P_1 +\lambda P_2$ is also a Poisson bivector for any $\lambda\in\mathbb R$.

Instead of using a basis of bivectors $P_{1,2}$, we can also use a frame of three invariant bivectors
\bq\label{eul-pi}
P'_i=\partial_i\wedge \mathcal L_X \partial_i\,,\qquad\mbox{where}\qquad \partial_i=\frac{\partial}{\partial M_i}\,,\qquad i=1..3\,,
\eq
related by the equation
\[
b_1P'_1+b_2P'_2+b_3P'_3=0\,.
\]
Since the Schouten-Nijenhuis bracket (\ref{br-sn}) between these bivectors is zero
\[
[\![P'_i,P'_j]\!]=0\,,\qquad i,j=1,2,3,
\]
we have three compatible to each other Poisson bivectors. The vector field $X$ (\ref{X-eul}) is Hamiltonian with respect to any of these invariant Poisson bivectors
 \[
 X=\frac{1}{a_i}P'_idf_1\,,\qquad X=\frac{1}{2}P'_idf_2\,,\qquad i=1,2,3.
 \]
 
 Let us now describe a result of our mathematical experiment. If we substitute (1.12) into (1.2) and solve the resulting equations
we get all these multivector tensor invariants and their combinations.
Furthermore we found  another invariant tensor field $T$ of type $(3,0)$ whose components are homogeneous cubic polynomials (\ref{cub-anz}) on $M_i$. This solution depends on 4 parameters, cannot be represented as a sum of absolutely antisymmetric and absolutely symmetric parts and is uniquely determined by the fact that multiplying $T$ by the invariant 1-forms $df_1$ and $df_2$ gives a single invariant symmetric tensor of type $(2,0)$
  \[P_s=X\vee X\] 
up to a factor 
 \[
 P_s^{ij}=c_{11}\sum_{k=1}^n T^{ijk}\frac{\partial f_1}{\partial M_k}=  c_{12} \sum_{k=1}^n T^{ikj}\frac{\partial f_1}{\partial M_k}=c_{13}\sum_{k=1}^n T^{kij}\frac{\partial f_1}{\partial M_k}
 \]
  and
\[
 P_s^{ij}=c_{21}\sum_{k=1}^n T^{ijk}\frac{\partial f_2}{\partial M_k}=  c_{22} \sum_{k=1}^n T^{ikj}\frac{\partial f_2}{\partial M_k}=c_{23}\sum_{k=1}^n T^{kij}\frac{\partial f_2}{\partial M_k}
 \]  
 where
 \[
 c_{11}+c_{12}+c_{13}=c_{21}+c_{22}+c_{23}=0\,,\qquad c_{ij}\in\mathbb R\,.
 \]
 The relation $T \left(df_1\wedge df_2\right)=0$ is valid as well. For the sake of brevity, we will not write all the components of this invariant tensor explicitly.

In the space of tensor fields of type (1,1), the only solution we obtained in the experiment has the form 
\[ (N)^i_j=g(f_1,f_2)\,\delta^{i}_{j}\,,
\] 
where $g(f_1,f_2)$ is arbitrary function on $f_1$ and $f_2$. Invariant mixed tensors of type (2,1) and (1,2) are combinations of the invariant tensor fields considered above.

\section{The Lagrange  case}
The equations of motion  are 
\[
\dot{\gamma}_1=2(\gamma_3M_2 -\gamma_2M_3)\,,\qquad \dot{\gamma}_2=2(\gamma_1M_3-\gamma_3M_1)\,,\qquad \dot{\gamma}_3=2(\gamma_2M_1 - \gamma_1M_2)\,,\]
 and
 \[
\dot{M}_1=-a\gamma_2\,,\qquad \dot{M}_2=a\gamma_1\,,\qquad \dot{M}_3=0\,,
\]
where $\gamma_1,\gamma_2,\gamma_3$ and $M_1,M_2,M_3$ are the coordinates on the phase space of this autonomous system of differential equations, $a$ is an arbitrary real parameter. A discussion of the equations, the physical meaning of the coordinates and the parameter $a$ can be found, for example, in the textbook \cite{bm}.

The divergence of the corresponding vector field $X$ is zero, so that the phase flow preserves the volume form $\Omega$ (\ref{vol-inv}) and the fully antisymmetric unit tensor $\mathcal E$ (\ref{lc-t}). We choose the following polynomials as independent scalar invariants
\[f_1=\gamma_1^2+\gamma_2^2+\gamma_3^2\,,\qquad f_2=\gamma_1M_1+\gamma_2M_2+\gamma_3M_3\,,\qquad
f_3=M_1^2+M_2^2+M_3^2+a\gamma_3\qquad\mbox{and}\qquad f_4=M_3\,,\]
which give rise to independent invariant differential 1-forms (\ref{df-1}).
\begin{prop}
In the Lagrange case, all solutions of the invariance equation in the space of differential forms of type $(0,2)$ and $(0,3)$ with components (\ref{cub-anz}) are tensor products of invariant 1-forms
\[
 \omega_2=dg_1(f_1,\ldots,f_4) dg_2(f_1,\ldots,f_4)\qquad \mbox{and}\qquad
 \omega_3=dg_1(f_1,\ldots,f_4) dg_2(f_1,\ldots,f_4)dg_3(f_1,\ldots,f_4)\,,\]
 where $g_k$ are arbitrary functions on first integrals.
\end{prop}
The proof consists in the explicit solution of the invariance equation by substitution (\ref{cub-anz}) and the analysis of the solutions thus obtained.

The experimental fact that there are no other solutions in the space of differential forms for this and other integrable systems is one of the reasons for the limited applicability of the mathematical theory of Poincar\'{e}-Cartan integral invariants in classical mechanics.

In contrast to the space of differential forms, in the space of multivector fields of type $(n, 0)$ the set of solutions of the invariance equation is not exhausted by tensor operations on a small set of basic invariants. Indeed, in the space of $(1,0)$ tensor fields, the general solution of the invariance equation is
\[Z=g_1(f_1,f_2,f_3,f_4)X+g_2(f_1,f_2,f_3,f_4)Y\,,\]
where
\bq\label{lag-y}
Y=-\gamma_2\partial_1+\gamma_1\partial_2-M_2\partial_4+M_1\partial_5\qquad\mbox{and}\qquad \partial_i=\frac{\partial}{\partial \gamma_i}\,,\quad \partial_{i+3}=\frac{\partial}{\partial M_i}\,,\quad i=1,2,3\,.
\eq
Moreover, in the space of vector fields there exists a conditional invariant $\partial_6$ associated with the linear first integral $f_4=M_3$ and the corresponding Noether symmetry
\[ L_X e_6\neq 0\,,\qquad\mbox{but}\qquad  (L_X e_6)df_4=0\,.\] 
In the following we will use this symmetry-related conditional invariant for the construction of other conditional invariants.

\subsection{Invariant tensor fields of type (2,0)}
Substituting a tensor field $P$ of type $(2,0)$ with components of the form (\ref{cub-anz}) into the equation $\mathcal L_XP=0$ and equating the coefficients with different degrees of the variables $\gamma$ and $M$ to zero, we obtain a system of 6860 algebraic equations for 3024 parameters. 

Solving these equations using computer algebra systems we obtain the absolute tensor invariants, which are divided by symmetric and antisymmetric tensor fields, among the latter there are fifty-one Poisson bivectors for which the Schouten-Nijenhuis bracket is zero, among which we can also distinguish six Poisson bivectors compatible with the canonical Poisson bivector $P_c$ on the Lie-Poisson algebra $e^*(3)$.

Following the \cite{r82, med96, ts08}, we take the following basis of solutions with  linear in variables $(\gamma,M)$ entries 
\begin{align}
P_1=&\left(\begin{matrix}
0& -M_3& M_2& 0& -\frac{a}2& 0\\M_3& 0& -M_1& \frac{a}2& 0& 0\\-M_2& M_1& 0& 0& 0& 0\\0& -\frac{a}2& 0& 0& 0& 0\\ \frac{a}2& 0& 0& 0& 0& 0\\0& 0& 0& 0& 0& 0
\end{matrix}\right)\,,\qquad
P_2=\left(\begin{matrix}
0& 2\gamma_3& -2\gamma_2& 0& 0& 0\\-2\gamma_3& 0& 2\gamma_1& 0& 0& 0\\2\gamma_2& -2\gamma_1& 0& 0& 0& 0\\0& 0& 0& 0& -a& 0\\0& 0& 0& a& 0& 0\\0& 0& 0& 0& 0& 0
\end{matrix}\right)\,,
\nn\\ 
\label{lag-lin}\\
P_3=&P_c=\left(
      \begin{matrix}
        0 &0 & 0 & 0 &\gamma_3 &-\gamma_2 \\
        0 & 0 & 0 & -\gamma_3 & 0 & \gamma_1 \\
        0 & 0 &0  & \gamma_2 & -\gamma_1 & 0 \\
        0 & \gamma_3 & -\gamma_2 & 0 & M_3 & -M_2 \\
        -\gamma_3 & 0 & \gamma_1 & -M_3 & 0 & M_1 \\
        \gamma_2 & -\gamma_1 & 0 & M_2 & -M_1 & 0 \\
      \end{matrix}
    \right)\,,\nn
\end{align}
where $P_3=P_c$ is the canonical Poisson bivector which defines the standard Lie-Poisson bracket on the algebra $e^*(3)$
\[
\{\gamma_i,\gamma_j\}=0\,,\qquad \{\gamma_i,M_j\}=\varepsilon_{ijk}\gamma_k\,,\qquad \{M_i,M_j\}=\varepsilon_{ijk}M_k\,,
\]
where $\varepsilon_{ijk}$ is the Levy-Civita tensor \cite{bm}.  

The vector field $X$ is written in multi-Hamiltonian form because these bivectors are compatible $[\![P_i,P_j]\!]=0$, $i,j=1,2,3$
\[
X=P_idf_i\,,\qquad i=1,2,3\,.
\]
Equality to zero of the Schouten-Nijenhuis brackets can be checked using the following representation 
\begin{align}
P_1&=\frac{1}2\left(\partial_1\wedge\mathcal L_{Z}\partial_1+\partial_2\wedge \mathcal L_{Z}\partial_2\right)\,,
\qquad
P_2=\frac{1}{2}\sum_{i=1}^3 \partial_i\wedge\mathcal L_X\partial_{i+3}+a\partial_{4}\wedge \partial_{5}\,,
\nn
\\
P_3&=\frac{1}{2}\left( \partial_4\wedge L_{Z}\partial_4+\partial_5\wedge \mathcal L_{Z}\partial_5\right)+\partial_6\wedge\mathcal L_{f_4Y} \partial_6\,,\quad\mbox{where}\quad Z=X-f_4Y\,,
\nn
\end{align}
These solutions of the invariance equation have been obtained and studied in \cite{r82, med96, ts08}. Let us proceed to the description of the previously unknown solutions.
 \begin{prop}
In the space of Poisson bivectors with components that are inhomogeneous polynomials of second order on the variables $(\gamma,M)$, there are two independent solutions of the invariance equation
\begin{align}
P_4= &\left(\begin{smallmatrix}
0&  -\gamma_1M_1 -\gamma_2 M_2&  -\gamma_2M_3&  M_1M_2&  -M_1^2&  -a\gamma_2\\
\gamma_1M_1 +\gamma_2M_2&  0&  \gamma_1M_3&  M_2^2&  -M_1M_2&  a\gamma_1\\
\gamma_2M_3&  -\gamma_1M_3&  0&  M_2M_3&  -M_1M_3&  0\\
-M_1M_2&  -M_2^2&  -M_2M_3&  0&  0&  -aM_2\\
M_1^2& M_1 M_2&  M_1M_3&  0&  0&  aM_1\\
a\gamma_2&  -a\gamma_1&  0&  aM_2&  -aM_1&  0\\
\end{smallmatrix}\right)\,,
\nn\\
\label{p45-lag}\\
P_5=&\left(\begin{smallmatrix}
0&f_1& 0& -\gamma_1M_2&\gamma_1 M_1& \gamma_2M_3-\gamma_3M_2\\
-f_1& 0& 0& -\gamma_2M_2& \gamma_2M_1& \gamma_3M_1 -\gamma_1 M_3\\
0& 0& 0& -\gamma_3M_2& \gamma_3M_1& \gamma_1M_2-\gamma_2M_1\\
\gamma_1M_2& \gamma_2M_2& \gamma_3M_2& 0& -\frac{a\gamma_3}{2}&\frac{a\gamma_2}{2}\\
-\gamma_1M_1& -\gamma_2M_1& \gamma_3M_1& \frac{a\gamma_3}{2}& 0& -\frac{a\gamma_1}{2}\\
\gamma_3M_2 -\gamma_2 M_3& \gamma_1M_3-\gamma_3M_1& \gamma_2M_1 -\gamma_1M_2& -\frac{a\gamma_2}{2}&\frac{a\gamma_1}{2}& 0\\
\end{smallmatrix}\right)
\nn
\end{align}
so that
\[
P_4 df_1 = 2f_2Y\,,\qquad P_4df_2=f_3Y\,,\qquad P_4df_3=3af_4Y\,,\qquad P_4df_4=af_4Y
\]
and
\[
P_5df_1=-2f_1Y\,,\qquad P_2df_2=-f_2Y\,,\qquad P_5df_3=-f_4X\,,\qquad P_5df_4=-\frac{1}{2}X\,.
\]
The Poisson invariant bivectors $P_4$ and $P_5$ are not compatible with the Poisson bivectors $P_1,P_2$ and $P_3$ (\ref{lag-lin}).
\end{prop}
In the space of symmetric tensor fields with cubic on variables $(\gamma,M)$ components, besides the tensor product of vector fields $X$ and $Y$, there are 3 more solutions of the invariance equation of rank one, two and four.For example, let us explicitly write the fourth rank solution
\[P_6=X\vee\Bigl( \sum_{i=1}^3 M_i\partial_i+a\partial_6  \Bigr)-aY\vee Z\,,
\] 
where $\vee$ is a symmetrized tensor product, $Y$ is a vector field (\ref{lag-y}) invariant with respect to the action of the initial vector field $X$, and the third vector field included in this definition is
\[
Z=\sum_{i=1}^3\Bigl(2 \gamma_i\partial_i+M_i\partial_{i+3}\Bigr)\,.
\]
It satisfies to the equation 
\[ \mathcal L_X Z=-aX\,.\]
The products of the symmetric tensor invariant $P_6$ with the invariant differential 1-forms $df_i$, $i=1,\ldots,4$ are linear combinations of the invariant vector fields $X$ and $Y$
\[P_6 df_1= f_2 X- 2af_1Y\,,\qquad P_6 df_2 = \frac{f_3}{2}\,X -\frac{3af_2}{2}\,Y\,,\]
and
\[P_6 df_3= \frac{3af_4}{2}\,X - af_3Y\,,\qquad 
P_6df_4=  \frac{a}{2} X - \frac{af_4}{2}\,Y\,.\]
According to Poincar\'{e} \cite{poi} we can use symmetric tensor fields to study dynamical systems. They are equivalent to antisymmetric tensor invariants in the theory of invariants.

The divergence of the vector field $X$ (\ref{z-div}) is zero and, therefore, the completely antisymmetric unit tensor $\mathcal E$ (\ref{lc-t}) is a solution of the invariance equation (\ref{g-inv}). Its product with the outer product of the invariant 1-forms $df_i$
\[P_7=\mathcal E df_1\wedge df_2\wedge df_3\wedge df_4\]
is invariant Poisson bivector of rank two such that $P_7df_i=0$\,, $i=1,\ldots,4$.

\subsection{Invariant tensor fields of type (3,0)}
The tensor product of absolute invariants of type $(2,0)$ and $(1,0)$ is an absolute tensor invariant of type $(3,0)$. The question naturally arises whether there are other absolute invariants whose components are also inhomogeneous cubic polynomials on the variables $(\gamma,M)$.
\begin{prop}
The invariance equation $\mathcal L_XT=0$ has a solution of the form
\begin{align}
T=&\tfrac{2(af_1 - f_2 M_3)}{a^2} \,\partial_1\wedge \partial_2\wedge \partial_3 -\tfrac{M_3}2\, \partial_4\wedge \partial_5\wedge \partial_6 \nn \\
 - &\partial_1\wedge \partial_2\wedge \left(\tfrac{2 M_3 (M_1 M_3 - a \gamma_1)}{a^2} \, \partial_4  - \tfrac{2 M_3 (M_2 M_3 - a \gamma_2)}{a^2}   \, \partial_5
  +\tfrac{ (M_1 \gamma_1 + M_2 \gamma_2 + 3 M_3 \gamma_3) a - 2 M_3^3}{a^2}  \, \partial_6\right) \nn\\
  +& \partial_1\wedge \partial_3\wedge\left(\tfrac{2 M_3 M_1 M_2}{a^2}\, \partial_4 -\tfrac{(M_1 \gamma_1 + M_2 \gamma_2 - M_3 \gamma_3) a + 2 M_3 M_2^2}{a^2}\, \partial_5 + \tfrac{2 M_3 (M_2 M_3 - a \gamma_2)}{a^2}\,   \partial_6\right) \nn\\ 
  +&\partial_1\wedge \partial_4\wedge\left( \tfrac{(M_1 M_3 -a \gamma_1)}{a}\, \partial_5  -\tfrac{M_1 M_2}{a}\, \partial_6 \right)
   + \tfrac{M_1^2 - M_3^2 + a \gamma_3}{a} \, \partial_1\wedge \partial_5\wedge \partial_6\label{lag-t3}\\
   +& \partial_2\wedge \partial_3\wedge\left(  \tfrac{(M_1 \gamma_1 + M_2 \gamma_2 - M_3 \gamma_3) a - 2 M_3 M_1^2}{a^2} \, \partial_4
    - \tfrac{2 M_3 M_1 M_2}{a^2}\, \partial_5 - \tfrac{2 M_3 (M_1 M_3 - a \gamma_1)}{a^2}\,\partial_6 \right)\nn\\
  +& \partial_2\wedge \partial_4\wedge\left(\tfrac{ M_2 M_3 - a\gamma_2}{a} \,\partial_5  - \tfrac{(M_2^2 - M_3^2 + a \gamma_3)}{a} \, \partial_6 \right)
   +\tfrac{M_1 M_2}{a}\, \partial_2\wedge \partial_5\wedge \partial_6\nn\\
    -&  \partial_3\wedge \partial_4\left( \gamma_3\, \partial_5  -\tfrac{ M_2 M_3-a \gamma_2}{a}\, \partial_6\right)  +\tfrac{M_1 M_3 -a \gamma_1}{a} \, \partial_3\wedge \partial_5\wedge \partial_6 \nn
\end{align}
for which
\[
X=Tdf_3\wedge df_4\,,
\]
and the products of the tensor field $T$ of type $(3,0)$ with other invariant 2-forms $df_i\wedge df_j$ are decomposed by the invariant vector fields $X$ and $Y$ as follows
\begin{align*}
Tdf_1\wedge df_2=&2\left(\frac{f_2f_4}{a^2} - \frac{f_1}{a}\right)X + \frac{4f_1f_4}{a}Y\,,\qquad
&Tdf_1\wedge df_4=\frac{2f_4^2}{a^2}X -  \frac{2H_2}{a}Y\,,\\ \\
Tdf_1\wedge df_3=&2\left( \frac{f_3f_4}{a^2}-\frac{f_2}{a}\right)X - 4\left(f_1 +\frac{f_2f_4}{a}\right)Y\,,\qquad
&Tdf_2\wedge df_3 = \frac{3f_4^2}{a}Xa -3f_2 Y\,,\\ \\
Tdf_2\wedge df_4=& \frac{3f_4}{2a}X - \frac{f_3-f_4^2}{a}Y\,. \qquad
\end{align*}
The components of the invariant bivectors $Tdf_i$, $i=1,\ldots,4$ are inhomogeneous polynomials of degree four and three on the variables $(\gamma,M)$, and the bivectors themselves do not satisfy the Jacobi identity.
\end{prop}
The proof consists of a direct verification of the above equations.

In the space of tensor fields of type $(3,0)$ with numerical components the invariance equation $\mathcal L_X T=0$ has no solutions. Let us therefore proceed to the equation (\ref{ug-inv2}) which in our case looks like
\bq\label{g-inv-x}
\bigl(\mathcal L_X T\bigr)dH=0\,,\qquad H=\sum_{k=1}^4e_k f_k\,,\qquad TdH\neq 0\,,\qquad e_k\in\mathbb R\,.
\eq
Here $H$ is the linear combination of the first integrals. The real parameters $e_k$ are found by solving this equation.
\begin{prop}
In the Lagrange case in the space of tensor fields of type $(3,0)$ with constant components, there are only three solutions of equation (\ref{g-inv-x})
\begin{align}
T_1=&\partial_1\wedge \partial_2\wedge \partial_6+\partial_2\wedge \partial_3\wedge \partial_4+\partial_3\wedge \partial_2\wedge \partial_5=\sum_{ijk=1}^3 \varepsilon_{ijk} \partial_i\wedge \partial_{j}\wedge \partial_{k+3}\,,
\nn\\
T_2=&\partial_1\wedge\partial_2\wedge\partial_3-a\partial_4\wedge\partial_5\wedge\partial_6=\sum_{ijk=1}^3 \varepsilon_{ijk}\left(\partial_j\wedge \partial_j\wedge \partial_k -a\partial_{i+3}\wedge \partial_{j+3}\wedge \partial_{k+3}\right)\,,
\label{lag-3t}\\
T_3=&\partial_1\wedge \partial_5\wedge \partial_6+\partial_2\wedge \partial_6\wedge \partial_3+\partial_3\wedge \partial_4\wedge \partial_5=\sum_{ijk=1}^3 \varepsilon_{ijk} \partial_i\wedge \partial_{j+3}\wedge \partial_{k+3}\,,
\nn
\end{align}
which give rise to Poisson invariant bivectors
\[
P_1=-\frac{1}{2}T_1df_3\,,\qquad P_2=T_2(df_1+df_4)\,,\qquad P_3=T_3df_2\,,
\]
and satisfy equations of the form (\ref{ug-inv2})
\[
\bigr(\mathcal L_X T_1\bigl)df_1=0\,,\qquad
\bigr(\mathcal L_X T_2\bigl)(df_1+df_4)=0\,,\qquad
\bigr(\mathcal L_X T_3\bigl)df_2=0\,.
\]
so that
\[
X=\frac{1}{2}T_1df_1\wedge df_3=T_2(df_1+df_4)\wedge df_2=T_3df_2\wedge df_3\,.
\]
\end{prop}
The proof consists in solving a system of algebraic equations using various computer algebra systems. The question of constructing conditional numerical invariants $T_i$ of type $(3,0)$ from an absolute numerical invariant $\mathcal E$ of type $(6,0)$ is open.

Conditional invariants $T_i$ (\ref{lag-3t}) are solutions of the equation (\ref{g-inv-x}) and are constructed without the participation of the second vector field $Y$ for the action of which they are absolute invariants $\mathcal L_YT_i=0$, $i=1,2,3$.

The tensor product of an absolute invariant of type $(2,0)$ and a conditional invariant of type $(1,0)$ is a conditional tensor invariant of type $(3,0)$ according to the Leibniz rule of differentiation. As an example, consider the tensor product of 
\[
T=P_i \partial_6\,,\qquad \partial_6=\frac{\partial}{\partial M_3}\,,
\] 
where $P_i$ is one of the absolute invariants of type $(2,0)$ found earlier, whose components are inhomogeneous polynomials on the variables $(\gamma,M)$.  The conditional invariant $T$ of type $(3,0)$ obtained in this way satisfies the relations
\[P_i=Tdf_4\quad\mbox{and}\quad (L_X T)\neq 0\,,\quad (L_X T)df_4=0\,.\]
As another example, consider a tensor field of type $(4,0)$ which is represented as 
\[
T=\Bigl(\partial_1\wedge\partial_2\wedge \partial_3\Bigr)\,\partial _6+\hat{T}\,,
\]
where $\mathcal E=\partial_1\wedge\partial_2\wedge \partial_3$ is a Levi-Civita tensor of type $(3,0)$, and in the tensor field $\hat T$ only two components are nonzero $\hat{T}^{4233}= 1$ and $\hat{T}^{5133} = -1$.  In this case
\[
X=\frac{1}{2f_4}\,Tdf_1df_2df_3\quad\mbox{and}\quad \mathcal L_XT\neq 0\,,\qquad \bigl(\mathcal L_X T\bigr)df_4=0\,.
\]
However, tensor products $T$ with $df_i$ or $\omega_{i,j}=df_i\wedge df_j$ will not satisfy any of the invariance conditions discussed. This example shows that to find conditional invariants in spaces of tensor fields of higher valence, the number of imposed conditional invariance equations must be increased accordingly.

Thus, in the space of tensor fields of type $(3,0)$ with $(\gamma,M)$ components cubic in variables (\ref{cub-anz}) there exists a non-empty set of absolute and conditional tensor invariants.  The question of the properties of these invariants, their classification and construction using arbitrary basic invariants remains open.

\section{The case of Kovalevskaya}
Six equations of motion of a rigid body in the Kovalevskaya case
\[
\dot{\gamma}_1=2\gamma_3M_2 - 4\gamma_2M_3\,,\qquad \dot{\gamma}_2=-2\gamma_3M_1 + 4\gamma_1M_3\,,\qquad\dot{\gamma}_3=2\gamma_2M_1 - 2\gamma_1M_2\,,
\]
and
\[
\dot{M}_1=-2M_2M_3\,,\qquad \dot{M}_2=2M_1M_3 - a\gamma_3\,,\qquad \dot{M}_3=a\gamma_2
\]
possess  first integrals
\[f_1=\gamma_1^2+\gamma_2^2+\gamma_3^2\,, \qquad f_2=\gamma_1M_2+\gamma_2M_2+\gamma_3M_3\,,\qquad f_3=M_1^2 + M_2^2 + 2M_3^2 + a\gamma_1\,,\]
and
\[f_4=(M_1^2 - M_2^2 - a\gamma_1)^2 + (2M_1M_2 - a\gamma_2)^2\,.
\]
Since the divergence of the corresponding vector field $X$ is zero, the phase flow preserves the volume form $\Omega$ (\ref{vol-inv}) and, therefore, these equations are integrable by quadratures according to the Euler-Jacobi theorem \cite{bm}.

\begin{prop}
In the Kovalevskaya case, all solutions of the invariance equation in the space of differential forms of type $(0,2)$ and $(0.3)$ with components of the form (\ref{cub-anz}) are tensor products of invariant 1-forms
\[
 \omega_2=dg_1(f_1,\ldots,f_4) dg_2(f_1,\ldots,f_4)\qquad \mbox{and}\qquad
 \omega_3=dg_1(f_1,\ldots,f_4) dg_2(f_1,\ldots,f_4)dg_3(f_1,\ldots,f_4)\,.
 \] 
 where $g_k$ are arbitrary functions on the first integrals.
\end{prop}
The proof consists in the explicit solution of the invariance equation by substitution (\ref{cub-anz}) and the analysis of the solutions thus obtained.

The corresponding vector field $X$  is the Hamiltonian vector field
 \[X=P_cdf_3\,,
 \] 
 where $P_c=P_3$ (\ref{lag-lin}) is the canonical Poisson bivector on the Lie-Poisson algebra $e^*(3)$ \cite{bm}. The components of the second invariant vector field \[Y=P_cdf_4\] are inhomogeneous polynomials of degree four on the variables $(\gamma,M)$
\begin{align*}
Y=&4\gamma_3\Bigl((M_1^2 + M_2^2)M_2+a(\gamma_1M_2-\gamma_2M_1)\Bigr)\partial_1-4\gamma_3\Bigl((M_1^2 + M_2^2)M_1-a(\gamma_1M_1+\gamma_2M_2) \Bigr) \partial_2\\
+&4\Bigl( (M_1^2 + M_2^2)(\gamma_2M_1 -\gamma_1 M_2)-a(\gamma_1^2 + \gamma_2^2)M_2\Bigr)\partial_3\\ 
+ &2\Bigl(2(M_1^2 + M_2^2)M_2M_3 -2a \bigl((\gamma_2M_1 - \gamma_1M_2)M_3 +2\gamma_3M_1M_2\bigr) + a^2\gamma_2\gamma_3\Bigr)\partial_4\\ 
-  &2\Bigl((M_1^2 + J2^2)M_1M_3-a(2(\gamma_1M_1+\gamma_2M_2)M_3 - \gamma_3(M_1^2 - M_2^2)) +a^2\gamma_1\gamma_3\bigr)\partial_5\\ 
+ & 2a\Bigl(\gamma_2(M_1^2 - M_2^2) - 2\gamma_1M_1M_2\Bigr) \partial_6\,.
\end{align*}
It is known that the vector field $X$ is bi-Hamiltonian \cite{mar98,ver14} and therefore we know several absolute invariants of the vector field $X$ in the Poisson bivector space. The following solutions of the invariance equation are new and cannot be found in the existing literature.

\subsection{Invariant tensor fields of type (2,0)}
Substituting the tensor field $P$ of type $(2,0)$ with components of the form (\ref{cub-anz}) into the equation $\mathcal L_XP=0$ and equating the coefficients with different degrees of the variables $\gamma$ and $M$ to zero, we obtain a system of 7164 algebraic equations for 3024 parameters. 

Solving the equations obtained in this way with the help of computer algebra systems, we obtain absolute tensor invariants, among which, contrary to the Lagrange case, there are no symmetric tensor fields. Among the symmetric bivectors obtained, only five satisfy the Jacobi identity, and three of them are compatible with the canonical Poisson bivector $P_c$ on the Lie-Poisson algebra $e^*(3)$.

The first of the invariant Poisson bivectors  is expressed by the first integrals and the canonical bivector $P_c=P_3$ (\ref{lag-lin})  
\[
P_1=(c_1f_1+ c_2 f_2+c_3)P_c\,,\qquad c_i\in\mathbb R\,,
\]
The components of the second invariant  Poisson bivector  have the form
\begin{align*}
P_{2}^{12}=& 2(f_2\gamma_3-f_1M_3)\,,\quad P_{2}^{13}= f_1M_2- 2f_2\gamma_2\,,\quad P_{2}^{1,4}=\gamma_1M_2 -\gamma_2 M_1)M_3+\tfrac{a\gamma_2 \gamma_3}{2}\,,
\\ \\
P_{2}^{15}=&\tfrac{\gamma_2f_3}{2} - f_2M_3\,,\quad P_{2}^{16}=(\gamma_3M_2 -\gamma_2M_3)M_3 -\tfrac{(f_3-a\gamma_1)\gamma_2}{2}\,,\quad
P_{2}^{23}=2f_2\gamma_1-f_1M_1\,,
\\ \\
P_{2}^{24}=&f_2M_3- \tfrac{f_3\gamma_3}{2}\,,\quad P_{2}^{25}=(\gamma_1M_2 - \gamma_2M_1)M_3+\tfrac{a\gamma_2\gamma_3}{2}\,,\quad
P_{2}^{26}=\tfrac{f_3\gamma_1}{2}+(\gamma_1M_3 -\gamma_3 M_1)M_3-\tfrac{a\gamma_2^2}{2}\,,
\\ \\
P_{2}^{34}=&\tfrac{(M_1^2 + M_2^2)\gamma_2}{2} +\gamma_3M_2M_3\,,\quad P_{2}^{35}=(\gamma_1M_3 -\gamma_3 M_1)M_3-\tfrac{f_3\gamma_1}{2}+\tfrac{a(\gamma_1^2 + \gamma_3^2)}{2}\,,
\\ \\
P_{2}^{36}=&(\gamma_2M_1 - \gamma_1M_2)M_3-\tfrac{a\gamma_2\gamma_3}{2}\,,\quad P_{2}^{45}=\tfrac{a\gamma_3M_1 - (M_1^2 + M_2^2)M_3}{2}\,,\quad
P_{2}^{46}=-\tfrac{f_3M_2}{2}\,,\quad P_{2}^{56}=\tfrac{f_3M_1}{2}  -f_2\,.
\end{align*}
This Poisson bivector is compatible with the canonical bivector $P_c$ and it satisfies the relations 
\[ P_{2} df_1= f_1X\,,\qquad P_{2}df_2=f_2X\,,\qquad P_{2}df_3 = f_3X + \frac{Y}{4}\,,\qquad P_{2}df_4=f_4X\,,\qquad P_{2}df_4=f_4X\,,\] 
from which follow expressions for the Casimir function
\[{C}_1=\frac{f_2}{f_1}\qquad\mbox{and}\qquad C_2=\frac{f_4}{f_1}\,,\qquad\mbox{so that}\qquad P_{2}d{C}_{1,2}=0\,.\]
%
%
%
\par\noindent
The components of the third Poisson bivector are equal to
\begin{align*}
P_{3}^{12}=&-2(\gamma_1^2 + \gamma_2^2)M_3\,,
\qquad
P_{3}^{13}=f_1M_2 - 2\gamma_2\gamma_3M_3\,,
\\ \\
P_{3}^{14}=&(\gamma_3M_2 -\gamma_1 M_3)M_1 +\gamma_1M_2M_3 -\frac{a\gamma_2\gamma_3}{2}\,,
\qquad
P_{3}^{15}= \gamma_3(M_2^2 + 2M_3^2) -f_2M_3+a\gamma_1\gamma_3\,,
\\ \\
P_{3}^{15}=&-\gamma_2M_3^2 -\frac{a\gamma_1\gamma_2}{2}\,,\quad
P_{3}^{23}=-f_1M_1 + 2\gamma_1\gamma_3M_3\,,
\\ \\
P_{3}^{24}=&f_2M_3-(M_1^2 - 2M_3^2)\gamma_3\,,\quad
P_{3}^{25}=\gamma_1 M_2M_3-\gamma_3 M_1M_2 -\gamma_2M_1 M_3  +\frac{a\gamma_2\gamma_3}{2}\,,
\\ \\
P_{3}^{26}=&\gamma_1M_3^2 -\frac{a\gamma_2^2}{2}\,,
\qquad
P_{3}^{34}=\frac{\gamma_2(2M_1^2 + 2M_3^2 + a\gamma_1)}{2}-(\gamma_1M_1 +\gamma_3M_3)M_2 \,,
\\ \\
P_{3}^{35}=&(\gamma_2M_2 -\gamma_3 M_3)M_1  - \gamma_1(M_2^2 + M_3^2)-\frac{a(\gamma_1^2 - \gamma_3^2)}{2}\,,
\qquad
P_{3}^{36}=-\frac{a\gamma_2\gamma_3}{2}\,,
\\ \\
 P_{3}^{4,5}=&M_3^3 + \frac{a\gamma_1M_3}{2}\,,
\quad
P_{3}^{46}=-M_2M_3^2 -\frac{ a\gamma_2M_1}{2}\,,\quad P_{3}^{56}=M_1M_3^2 +\frac{ a\gamma_1M_1}{2}\,.
\end{align*}
This Poisson bivector is also compatible with the canonical bivector $P_c$. 

Multiplying this bivector by the 1-form $df_1,\ldots,df_4$ we obtain
\[
P_{3} df_1=f_1X\,,\qquad P_{3}df_2=\frac{1}{2}f_2X\,,\qquad
P_{3} df_3= \frac{1}{2}f_3X + \frac{1}{4}Y\,,\qquad P_{3}df_4=f_4X -\frac{1}{2}f_3Y\,.
\]
As Casimir functions we can take \[{C}_1=\frac{f_2}{\sqrt{f_1}}\qquad\mbox{and}\qquad {C}_2=\frac{f_3^2-f_4}{f_1}{f_1}\,,\qquad\mbox{so that}\qquad P_{3}d{C}_{1,2}=0\,.\]
For completeness, we also give two invariant Poisson bivectors which are not compatible with the canonical bivector $P_c$. The components of the first one have the form
\begin{align*}
P_4^{12}=&2\gamma_3(\gamma_1M_1 + \gamma_2M_2)-4(\gamma_1^2 +\gamma_2^2)M_3\,,\quad P_4^{13}=2(\gamma_1^2 + \gamma_3^2)M_2-2\gamma_2 (\gamma_1M_1+\gamma_3M_3)\,,
\\ \\
P_4^{14}=&\gamma_3 M_1M_2 -2(\gamma_2M_1-\gamma_1M_2)M_3\,,\quad
P_4^{15}=\gamma_3(M_2^2 + a\gamma_1)-2 (\gamma_1M_1+\gamma_2M_2)M_3 \,,
\\ \\
P_4^{16}=&(\gamma_3M_2 - 2\gamma_2M_3)M_3-a\gamma_1\gamma_2\,,\quad 
P_4^{23}=2\gamma_1(\gamma_2M_2 + 2\gamma_3M_3)-2(\gamma_2^2 + \gamma_3^2)M_1\,,
\\ \\
P_4^{24}=& 2(\gamma_1M_1 +\gamma_2M_2)M_3-\gamma_3M_1^2\,,\quad P_4^{25}=2(\gamma_1M_2-\gamma_2M_1)M_3 - \gamma_3(M_1M_2 - a\gamma_2)\,,
\\ \\
P_4^{26}=&(2\gamma_1M_3-\gamma_3M_1)M_3-a\gamma_2^2\,,\quad P_4^{34}=2\gamma_3M_2M_3 + (\gamma_2M_1 -\gamma_1M_2)M_1\,,
\\ \\
P_4^{35}=&\gamma_3(a\gamma_3 - 2M_1M_3) + (\gamma_2M_1 -\gamma_1M_2)M_2\,,\quad P_4^{36}=(\gamma_2M_1 -\gamma_1M_2)-a\gamma_2\gamma_3\,,
\\ \\
P_4^{45}=&\tfrac{a\gamma_3M_1}{2} -(M_1^2 + M_2^2) M_3\,,\quad P_4^{46}= -\tfrac{a\gamma_2M_1}{2}-M_2M_3^2\,,\quad
P_4^{56}=M_1 M_3^2-\tfrac{a(\gamma_2M_2+\gamma_3M_3)}{2}\,.
\end{align*}
The second Poisson invariant bivector, which is incompatible with $P_c$, is expressed by $P_4$ and $P_c$.
\[P_5=P_4-\frac{f_3}{2}P_c\]
The rank of the Poisson bivectors $P_4$ and $P_5$ are two and four, respectively. In this case we have
\[
P_5df_1= 2f_1X\,,\qquad P_5df_2=\frac{3}{2}f_2X\,,\qquad P_5 dh3=\frac{1}{2}f_3X\,,\qquad 
P_5df_4= 2f_4X -\frac{f_3}{2}Y\,,
\]
and as Casimir functions of the bivector $P_5$ we can take the functions
\[
C_1=\frac{f_2^{2/3}}{f_1^{1/2}}\,,\qquad C_2=\frac{f_3^2}{f_1^{1/2}}\,,\qquad P_5dC_{1,2}=0\,.
\]
This completes the list of Poisson's invariant tensors that we have obtained in our mathematical experiment.
\begin{prop}
For the Kovalevskaya system, the set of solutions of the invariance equation \[\mathcal L_X\,P=0\] in the Poisson bivector space with components of the form (\ref{cub-anz}) consists of the above five bivectors $P_1,\ldots,P_5$.
\end{prop}
The proof consists of the direct solution of the equation of invariance using substitution (\ref{cub-anz}).

Since the divergence of the vector fields $X$ and $Y$ is zero, the tensor product of the completely antisymmetric unit tensor $\mathcal E$ (\ref{lc-t}) with the product of four invariant 1-forms $df_i$ is given by 
\[P_6=\mathcal E df_1\wedge df_2\wedge df_3\wedge df_4\] 
is an invariant second-rank Poisson bivector whose components are inhomogeneous polynomials of degree 6 on the variables $(\gamma,M)$.

\subsection{Invariant tensor fields of type (3,0)}
Substituting the tensor field $T$ of type $(3,0)$ with components of the form (\ref{cub-anz}) into the invariance equation (\ref{g-inv}) and equating the coefficients with different degrees of the variables $\gamma$ and $M$ to zero, we obtain a system of 43953 algebraic equations for 18144 unknowns. Solving the equations obtained in this way using computer algebra systems, we obtain the solution $T=P_cX$ and, contrary to the Lagrange case, there are no other solutions.

To get a counterpart of the solution (\ref{lag-t3}) for the Lagrange case we have to use in Kovalevskaya case the higher order polynomial anzats for the components of tensor field $T$ of type $(3,0)$. We will not present this solution for brevity.

Then, as in the Lagrange case, we proceed to the solution of the equation of relative invariance (\ref{g-inv-x}), which involves a linear combination of the first integrals $H=\sum_k e_kf_k$, in which the unknown coefficients $e_k\in \mathbb R$ are obtained in the process of solution.
  \begin{prop}
In the Kovalevskaya case in the space of tensor fields of type $(3,0)$ with constant components, there is only one solution
\[
T=\partial_1\wedge \partial_5\wedge \partial_6+\partial_2\wedge \partial_6\wedge \partial_4+\partial_3\wedge \partial_4\wedge \partial_5=\sum_{ijk=1}^3 \varepsilon_{ijk} \partial_i\wedge \partial_{j+3}\wedge \partial_{k+3}\,,
\]
relative invariance equations (\ref{g-inv-x})
\[
\bigl(\mathcal L_X T\bigr)dH=0\,,\qquad TdH\neq 0
\]
which satisfies the following relations
\[
X=Tdf_2\wedge df_3\,,\qquad \bigr(\mathcal L_X T\bigl)df_2=0\,,\qquad Tdf_2=P_c\,.
\]
\end{prop}
The proof consists of solving a system of algebraic equations using various computer algebra systems.

Thus, we proved that for the equations of motion of a rigid body in the Euler, Lagrange and Kovalevskaya cases there exists an antisymmetric numerical tensor field $T$ with a zero Schouten-Nijenhuis bracket $[\![T,T]\!]=0$, which allows us to rewrite the original vector field in the form of
\[
X=T\omega\,,\qquad\mbox{where}\qquad \mathcal L_x\omega=0\,, \qquad d\omega=0
\]
where $\omega$ is an invariant differential form of type $(0,2)$.

\section{Conclusion}
For a vector field $X$ without divergence, the completely antisymmetric unit tensor fields of valences $(0,n)$ and $(n,0)$ are basic tensor invariants. 
Using these tensor invariants and first integrals $f_1,\ldots,f_k$ we can construct a set of other tensor invariants.

Our mathematical experiment aimed to find all tensor invariants of the equations of motion of a rigid body in the cases of Euler, Lagrange and Kovalevskaya with the fixed form of the components (\ref{cub-anz}) and low valence 1,2 and 3.

The main result of the experiment is the fact that for Lagrange and Kovalevskaya systems in the space of differential forms all solutions of the invariance equation (\ref{g-inv}) obtained by us can be constructed from the basic invariant $\Omega$ and first integrals by differentiation and tensor multiplication. In the space of multivector fields, several absolute and conditional invariants have been found for these systems which, at least for the moment, cannot be constructed from the basic invariants by standard tensor operations. We emphasise again that we are talking about global tensor invariants uniquely defined on the whole phase space, since locally, in the neighbourhood of Liouville tori, there is no difference between the number of solutions in the spaces of covariant and contravariant tensor fields \cite{bog96}.

 The study was carried out with the financial support of the Ministry of Science and Higher Education of the Russian Federation in the framework of a scientific project under agreement No. 075-15-2024-631».

\end{document}